\documentclass{appolb}
\usepackage[pdftex]{graphicx}
\usepackage{amsmath}
\usepackage{amssymb}
\usepackage{euscript}
\usepackage{multirow}
\usepackage{color}
\usepackage{colortbl}
\usepackage{calc}
\usepackage{cite}

\makeatletter
\newlength\twolinebox@linelength
\newlength\twolinebox@columnheight
\newcommand{\twolinebox}[2]{%
   \setlength{\twolinebox@linelength}%
             {\maxof{\widthof{#1}}{\widthof{#2}}}%
   \setlength{\twolinebox@columnheight}{\heightof{#1}+\depthof{#1}+0.2em+0.4em/2+\heightof{0}/2}%
   \raisebox{0pt}[\twolinebox@columnheight][\heightof{\vbox{\vskip0.2em\hbox to 
   \twolinebox@linelength {#1\hfil}\vskip0.4em\hbox to 
   \twolinebox@linelength {#2\hfil}}}+\depthof{\vbox{\vskip0.2em\hbox to 
   \twolinebox@linelength {#1\hfil}\vskip0.4em\hbox to 
   \twolinebox@linelength {#2\hfil}}}-\twolinebox@columnheight+0.2em]{\vbox to 
   \twolinebox@columnheight{\vskip0.2em\hbox to 
   \twolinebox@linelength {#1\hfil}\vskip0.4em\hbox to 
   \twolinebox@linelength {#2\hfil}}}%
}
\newcommand\WINHAC[0] {\textsf{WINHAC}}
\newcommand\WINHACpp[0] {\textsf{WINHAC++}}
\newcommand\ZINHAC[0] {\textsf{ZINHAC}}
\newcommand\PYTHIA[0] {\textsf{PYTHIA}}
\newcommand\HORACE[0] {\textsf{HORACE}}
\newcommand\SANC[0]   {\textsf{SANC}}
\newcommand\HERWIG[0] {\textsf{HERWIG}}

\newcommand\MC@NLO[0] {\textsf{MC@NLO}}

\newcommand\PHOTOS[0] {\textsf{PHOTOS}}

\begin{document}
\eqsec  
\title{Drell--Yan processes with WINHAC%
\thanks{The work is partly supported  
by the Polish National Centre of Science grants DEC-2011/03/B/ST2/00220
and DEC-2012/04/M/ST2/00240,
and by the Programme of the French--Polish 
Cooperation between IN2P3 and COPIN no.\ 05-116.
\\
Presented by W.\ P\l{}aczek at the XXXVII International Conference
of Theoretical Physics ``Matter To The Deepest'',
Ustro\'n, Poland, 1--6 September 2013.
}%
}
\author{%
W.\ P\l{}aczek \vspace{-2mm}
 \address{Marian Smoluchowski Institute of Physics, Jagiellonian University,\\
         ul.\ Reymonta 4, 30-059 Krakow, Poland.}
\vspace{5mm} \\      
S.\ Jadach \vspace{-2mm}
\address{Institute of Nuclear Physics, Polish Academy of Sciences,\\
  ul.\ Radzikowskiego 152, 31-342 Krakow, Poland.}
\vspace{5mm} \\      
M.~W.\ Krasny \vspace{-2mm}
 \address{Laboratoire de Physique Nucl\'eaire et des Hautes \'Energies, \\
          Universit\'e Pierre et Marie Curie Paris 6, 
          Universit\'e Paris Diderot Paris 7, 
          CNRS--IN2P3,
          4 place Jussieu, 75252 Paris Cedex 05, France.}
}
\maketitle
\begin{abstract}
We present the Monte Carlo event generator \WINHAC\ for 
Drell--Yan processes in proton--proton, proton--antiproton, 
proton--ion and ion--ion collisions. 
It features multiphoton radiation
within the Yennie--Frautschi--Suura exclusive exponentiation scheme with
${\cal O}(\alpha)$ electroweak corrections for the charged-current 
($W^+/W^-$) processes
and multiphoton radiation generated by \PHOTOS\ for neutral-current
($Z+\gamma$) ones. 
For the initial-state QCD/QED parton shower and hadronisation 
it is interfaced with \PYTHIA. 
It includes several options, e.g.\ for the polarized $W$-boson production, 
generation of weighted/unweighted events, etc. 
\WINHAC\ was cross-checked numerically at the per-mille level 
with independent Monte Carlo programs, such as \HORACE\ and \SANC.
It has been used as a basic tool for developing and testing some new
methods of precise measurements of the Standard Model parameters at the LHC, 
in particular the $W$-boson mass.
Recently, it has been applied to simulations of double Drell--Yan processes 
resulting from double-parton scattering, 
in order to assess their influence on the Higgs-boson detection at the LHC
in its $ZZ$ and $W^+W^-$ decay channels.
\end{abstract}

\section{Introduction}
\label{introduction}

The Drell--Yan (DY) process, i.e. lepton-pair production in hadronic collisions, 
played an important role in the past in testing the parton model as well as 
the quantum chromodynamics (QCD) as the theory of strong interactions. 
While at the low-energy hadron 
colliders the lepton pairs were produced through the virtual $\gamma$ exchange,
in the recent high-energy colliders (Tevatron, LHC) the collision energy 
is sufficient to produce $W^{\pm}$ and $Z$ bosons. Moreover, the cross sections
for these processes turn out to be relatively high. Therefore, they
they can be used to improve experimental precision of some Standard Model (SM) 
parameters values, in particular the $W$-boson mass $M_W$ and width $\Gamma_W$,
the weak-mixing angle $\sin^2\theta_W$ and the strong coupling constant 
$\alpha_s$.
According to PDG~2012~\cite{Beringer:1900zz} the experimental errors on
the $W$-boson mass and width are respectively:
$\delta M_W = 15\,$MeV,
$\delta \Gamma_W = 42\,$MeV,
while while the corresponding errors for the $Z$ boson are 
$\delta M_Z = 2.1\,$MeV, $\delta \Gamma_Z = 2.3\,$MeV. 
A difference between the opposite-charge $W$ masses is measured even worse: 
$\delta (M_{W^+}-M_{W^-}) = 600\,$MeV.
Smaller errors of the $W$-boson mass and width will allow for a better indirect
determination of the SM Higgs-boson mass \cite{Schael:2013ita}. 
In the case of a direct Higgs discovery at the LHC, 
this will provide the important consistency test of the Standard Model
\cite{Schael:2013ita}. 
In order to match the precision of other SM parameters in such fits/tests, 
$M_W$ should be measured at the LHC with the accuracy of $10\,$MeV or 
better \cite{Haywood:1999qg,Stark:EPS-HEP2013}. 

In theoretical descriptions of the DY processes reaching a
sufficiently high precision for the LHC experiments 
requires including besides the QCD effects also the electroweak
(EW) corrections, in particular the QED final-state radiation (FSR) 
\cite{Baur:1998kt}. It is known that that including ${\cal O}(\alpha)$ 
EW corrections is not sufficient for the $W$-mass precision target at the LHC,
because the higher-order FSR effects can shift $M_W$ by $\sim 10\,$MeV,  
see e.g.~\cite{CarloniCalame:2003ux}.    
Regions of a high transverse momentum of a charged lepton, a large $W$-boson
transverse mass and a large $Z$-boson invariant mass 
are used at the LHC for various ``new physics'' searches.
In these regions the ${\cal O}(\alpha)$ EW corrections beyond FSR 
can be of the size of $20$--$30\%$ \cite{Dittmaier:2001ay,Baur:2004ig}. 
Because of high statistics of DY events expected at the LHC, 
these effects must be included in theoretical description of the SM background. 
However, as was argued in \cite{Placzek:2009jy}, for realistic 
predictions, the EW corrections should be combined in this case 
with the QCD parton-shower effects.
High statistics of DY data at the LHC will be important to reduce
uncertainties of parton distribution functions (PDFs) over a wide range
of the $(x,Q^2)$ domain, see e.g.~\cite{Forte:2013wc}.
DY processes are also treated at the LHC as the so-called 
`standard candle' processes (detector calibration, normalisation, etc.),
see e.g.~\cite{Krasny:2007cy}.
Last but not least, in addition to the standard single DY processes, 
the so-called double Drell--Yan processes (DDYP), being a product
of double-parton scattering (DPS), may play an important role at the LHC, 
in particular for the Higgs-boson searches in the $4l$ and $2l2\nu$ channels,
resulting from its decays into $ZZ$ and $W^+W^-$ pairs.
\cite{Krasny:2013aca}. 

Therefore, a precise theoretical description of DY processes is very important
for the LHC. 
In order to be fully exploited by the experiments, such a description 
ought to be provided in form of a Monte Carlo event generator.  
In this paper we briefly describe the Monte Carlo event generator 
\WINHAC~\cite{WINHAC:MC} dedicated to precise theoretical predictions 
for the DY processes.
In Section~2 we describe the implementation of the EW corrections  
as well the QCD effects and discuss their interplay, 
while in Section~3 we briefly review some applications of \WINHAC\ 
to the LHC physics. Finally, Section~4 contains summary and outlook.

\section{Electroweak corrections and QCD effects}
\label{sec:EWC}

In the current version of \WINHAC~\cite{WINHAC:MC} the description 
of the charged-current Drell--Yan (CC DY) processes is more advanced.
It features multiphoton radiation from all charged particles involved
in the hard process, implemented within the Yennie--Frautschi--Suura (YFS) 
exclusive exponentiation scheme \cite{YFS:1961} including the ${\cal O}(\alpha)$
EW corrections. In order to avoid ambiguities related to photon radiation
from light quarks, the QED initial-state radiation (ISR) is subtracted
from the EW corrections, however not in 
$\overline{\rm MS}$ or DIS schemes, as they are not well suited for a Monte Carlo
event generator. Instead we include in the program three options of the ISR
subtraction in a gauge-invariant way, with a default one called the YFS scheme,
in which from virtual EW corrections for the full CC DY process one subtracts
the YFS virtual form factor plus terms $\sim (1/2)Q_i^2[\ln(s/m_i^2) - 1]$,
where $Q_i$ is a quark electric charge in the units of the positron charge and
$m_i$ is its mass%
\footnote{A similar subtraction of QED corrections from the EW ones 
is done in Ref.~\cite{Wackeroth:1996}, however our results differ by some
constant terms.}. 
Generation of QED ISR effects is left to general-purpose parton shower
generators, such as \PYTHIA\ or \HERWIG, 
to which \WINHAC\ is, or will, be interfaced. 
In these generators, in addition to primary QCD effects, also QED ISR
can be generated using the parton shower algorithm. In such an approach 
QED radiation is intertwined with the dominant QCD radiation.
This solution is, in our opinion, better than the one in which photon radiation
from quarks is generated completely independently of the QCD effects.    
More details on the implementation of EW corrections for the CC DY processes 
in \WINHAC\ can be found in 
Refs.~\cite{Placzek2003zg,Placzek:2009jy,Bardin:2008fn}.

\WINHAC\ has been extensively tested numerically and cross-checked with
independent Monte Carlo programs. For the ${\cal O}(\alpha)$ and higher-order 
QED FSR effects is was compared with \HORACE~\cite{CarloniCalame:2004qw},
while for the ${\cal O}(\alpha)$ EW corrections is was compared with 
the \SANC\ Monte Carlo integrator \cite{Bardin:2008fn}. In both cases
the agreement at the per-mille level or better between the results of 
the programs was found. Currently, there is ongoing work within the LHC
Electroweak Working Group on detailed comparisons of various theoretical 
calculation and Monte Carlo programs for DY processes -- the results should 
be published soon.

For the NC DY process, in the current version of \WINHAC, only QED FSR effects are 
included through the \PHOTOS\ Monte Carlo generator. It implements multiphoton
radiation in particle decays using a leading-log-type iterative algorithm
where some important non-leading corrections are taken into account 
\cite{Golonka:2005pn}.
It was shown that in the case of $Z$-boson decays its predictions are
in a very good agreement with the ones of the ${\cal O}(\alpha^2)$ 
YFS exponentiation.

The \WINHAC\ program is a flexible Monte Carlo event generator. 
It includes several options allowing to choose between different collider types: 
proton--proton, proton-antiproton,
ion--ion, between various EW parameter schemes, initial-state quark flavours,
final-state lepton flavours, intermediate bosons, types of radiative
corrections, weighted or unweighted events, etc. Important options are
also possibilities to generate the CC DY processes with polarized $W$-bosons 
(transverse or longitudinal) in some predefined reference frames or
in a user-defined ones. 

For QCD effects the current version of \WINHAC\ is interfaced with the parton
shower generator \PYTHIA~{\sf 6.4}. It generates QCD/QED ISR using parton shower
algorithm, performs proton-remnant treatment, hadronisation and particle decays.
Two kind of interfaces are available now: the first one is the internal interface 
in which the \PYTHIA\ routines are called directly from the \WINHAC\ code and 
the second one is based on the Les Houches Accord in which events from 
\WINHAC\ are transmitted to \PYTHIA\ 
through special LHA-format files \cite{Alwall:2006yp}.   
The former interface is less universal -- it allows to use only some limited
set-up of the \PYTHIA\ parameters, but is faster and more flexible for some 
dedicated studies, and also includes corrections to {\sf PYTHIA~6} for its 
in improper predictions of lepton transverse momenta \cite{Krasny:2012pa}.
The second interface is more universal -- it allows for any set-up
of the \PYTHIA\ parameters, including \PYTHIA\ tunes. In this case Monte Carlo
events from \WINHAC\ are written in the LHA format into disk files and 
read in by \PYTHIA, which performs further processing of events. 
We have also added a possibility to transmit the hard-process
event from \PYTHIA\ back to \WINHAC, through another LHA file, 
for some data analysis. Instead of ordinary disk files we prefer to use the
UNIX named (FIFO) pipes, for which the input/output operations are 
identical as for ordinary files,
but the data transmission goes through RAM, not disk, and thus is much faster.
Moreover, one does not need to care about overloading disk space with huge
data files when high event statistics are generated.

\begin{table}[!ht]
\centering
{\scriptsize
\begin{tabular}{||c|c|c|c|c|c|c||}
\hline\hline
$p_T^{\mu}$ [GeV] & $>25$ & $>50$ & $>100$ & $>200$ & $>500$ & $>1000$ \\
\hline
\multicolumn{7}{||c||}{No QCD} \\
\hline
$\sigma_0$ [pb] & $4779.0\,(2)$ & $30.34\,(1)$ & $1.944\,(3)$ &
$0.178\,(1)$ & $0.0051\,(1)$ & $0.00015\,(1)$\\
\hline
$\delta_{\rm EW}$ [\%] & $-2.748\,(3)$ & $-6.10\,(3)$ & $-8.7\,(1)$ &
$-12.8\,(4)$ & $-21.4\,(1.3)$ & $-28\,(5)$\\
\hline
$\delta_{\rm weak}$ [\%]      & $-0.108\,(0)$ & $-1.193\,(1)$ & $-3.88\,(1)$
& $-7.36\,(4)$ & $-15.7\,(3)$ & $-23\,(2)$\\
\hline
\multicolumn{7}{||c||}{With QCD PS ({\PYTHIA})} \\
\hline
$\sigma_0$ [pb] & $4096.1\,(2)$ & $254.86\,(4)$ & $10.025\,(8)$ &
$0.683\,(2)$ & $0.0114\,(2)$ & $0.00024\,(2)$\\
\hline
$\delta_{\rm EW}$ [\%] & $-2.548\,(3)$ & $-4.99\,(1)$ & $-4.98\,(6)$ & 
$-7.1\,(2)$ & $-13.2\,(1.8)$ & $-20\,(11)$ \\
$\delta_{\rm weak}$ [\%]      & $-0.113\,(0)$ & $-0.250\,(0)$ & $-0.91\,(0)$
& $-2.16\,(1)$ & $-7.0\,(2)$ & $-15\,(2)$\\
\hline\hline
\end{tabular}
}
\caption{Born cross sections as well as electroweak and `weak' corrections without
(upper part) and with QCD effects from \PYTHIA\ (lower part)
for the $W^+ \rightarrow \mu^+\nu_{\mu}$
channel at the LHC collision energy of $14\,$TeV corresponding to the increasing
lower cut on the muon transverse momentum.}
\label{table1}
\end{table}

\begin{table}[!ht]
\centering
{\scriptsize
\begin{tabular}{||c|c|c|c|c|c|c||}
\hline\hline
$p_T^{\mu}$ [GeV] & $>25$ & $>50$ & $>100$ & $>200$ & $>500$ & $>1000$ \\
\hline
\multicolumn{7}{||c||}{No QCD} \\
\hline
$\sigma_0$ [pb] & $3720.1\,(1)$ & $22.45\,(1)$ & $1.211\,(2)$ &
$0.0971\,(4)$ & $0.00211\,(3)$ & $0.000052\,(2)$\\
\hline
$\delta_{\rm EW}$ [\%] & $-2.612\,(3)$ & $-6.16\,(3)$ & $-8.9\,(1)$ &
$-12.9\,(4)$ & $-21.6\,(1.5)$ & $-32\,(5)$\\
\hline
$\delta_{\rm weak}$ [\%]      & $-0.106\,(0)$ & $-1.094\,(1)$ & $-3.72\,(1)$
& $-7.13\,(4)$ & $-14.4\,(3)$ & $-22\,(2)$\\
\hline
\multicolumn{7}{||c||}{With QCD PS ({\PYTHIA})} \\
\hline
$\sigma_0$ [pb] & $3234.3\,(1)$ & $247.49\,(4)$ & $10.931\,(8)$ &
$0.832\,(2)$ & $0.0133\,(3)$ & $0.00027\,(4)$\\
\hline
$\delta_{\rm EW}$ [\%] & $-2.406\,(3)$ & $-4.85\,(1)$ & $-4.47\,(5)$ & 
$-5.5\,(2)$ & $-8.7\,(1.3)$ & $-15\,(8)$ \\
$\delta_{\rm weak}$ [\%]      & $-0.110\,(0)$ & $-0.210\,(0)$ & $-0.59\,(0)$
& $-1.10\,(1)$ & $-2.8\,(1)$ & $-5\,(1)$\\
\hline\hline
\end{tabular}
}
\caption{Born cross sections as well as electroweak and `weak' corrections without
(upper part) and with QCD effects from \PYTHIA\ (lower part)
for the $W^- \rightarrow \mu^-\bar{\nu}_{\mu}$
channel at the LHC collision energy of $14\,$TeV corresponding to the increasing
lower cut on the muon transverse momentum.}
\label{table2}
\end{table}
As was argued in Ref.~\cite{Placzek:2009jy}, in order to provide realistic
predictions of the EW effects in DY processes at the LHC and other hadron colliders,
the QCD parton shower effects must be taken into account. 
The QCD initial-state parton shower in DY processes modifies considerably
the transverse momenta of the final-state leptons. Since a lower cut on the 
charged-lepton transverse momentum $p_T^l$ is one of the primary cuts used by the
LHC experiments for the DY processes, this affects all the DY observables.
In Table~\ref{table1} we present the results for the Born cross sections as well as
the EW and `weak' corrections for the $W^+ \rightarrow \mu^+\nu_{\mu}$ channel
of the CC DY process at the LHC collision energy of $14\,$TeV 
for the increasing lower cut on $p_T^l$. 
In the upper part of the table we show the results without 
the QCD parton-shower effects, while in the lower part the ones in which
these effects, as generated by \PYTHIA, are included. 
In Table~\ref{table2} we present
similar results for the $W^- \rightarrow \mu^-\bar{\nu}_{\mu}$ channel.
The so-called `weak' corrections correspond to the difference of 
the EW corrections and a dominant gauge-invariant part of the QED
corrections, as defined in Ref.~\cite{Placzek:2009jy}.
One can see by comparing the upper and lower parts of the tables
that the QCD parton-shower corrections affect considerably not only 
the total cross sections but also both EW and `weak' corrections. 
These changes depend strongly 
on the $p_T^l$ cuts and are different for $W^+$ than for $W^-$.   
In particular, the `weak' corrections can be up to a factor $\sim 7$ smaller
in the presence of the QCD effects than without them. 
Therefore, any theoretical predictions of the EW corrections 
for the DY observables at the LHC without inclusion of the QCD parton-shower 
effects are not realistic. On the other hand, the EW corrections can be
quite sizeable, up to $\sim 20\%$, thus they also have to be taken into account.
One can conclude that any Monte Carlo event generator for precision 
physics at the LHC involving the DY processes should include both these effects.
This is particularly important for the $W$-boson mass measurement with the 
precision target below  $10\,$MeV. In this case even more important than 
the change of the size of the EW corrections after including the QCD effects 
is the change of their shape for the $p_T^l$ distribution, 
which is the main observable for the $M_W$ measurements at the LHC 
\cite{Placzek:Ustron2013}.

\section{Applications}
\label{sec:appl}

We have already applied \WINHAC\ to several dedicated physics studies
of possible measurements at the LHC where the DY processes may play 
an important role.   
Our first study was devoted to a possibility of the experimental investigation
of the electroweak symmetry breaking mechanism in proton--ion collisions at the LHC
by exploiting the effective beams of polarized $W$-bosons and their interactions
with spectator quarks \cite{Krasny:2005cb}.
Then we proposed a way to use the NC DY process, i.e. single-$Z$ production,
at the LHC as a `standard candle' for precision measurements of the SM parameters,
particularly the ones related to the $W$-boson physics \cite{Krasny:2007cy}.

Our main focus was on investigating the possibility to measure the $W$-boson mass
$M_W$ as well as the difference $\Delta M_{W^{\pm}} = M_{W^+} - M_{W^-}$ 
at the LHC with the precision
of $10\,$MeV or better. This was the subject of a series of papers
\cite{Fayette:2008wt,Krasny:2010vd}.
We discussed important differences between the DY processes
at the Tevatron and at the LHC and showed that the $M_W$ measurements
procedures used at the Tevatron cannot be applied at the LHC.
We proposed four bias-reducing observables to be used by the LHC experiments
in such measurements and argued that a dedicated fix-target ``LHC-support'' 
experiment with a high-intensity muon beam is needed to achieve the aimed 
$M_W$ precision target \cite{Krasny:2010vd}. 

Recently, using \WINHAC, we have performed studies of the double Drell--Yan 
processes (DDYP), resulting from double-parton scattering (DPS), and their
possible influence on the Higgs-boson searches in its $ZZ$ and $WW$ decay channels.
We have found that DDYP can produce an excess of events
in the Higgs-signal region corresponding to the Higgs mass of 
$\sim 125\,$GeV. This excess can contribute to the background to the Higgs
signal, and even explain, under certain conditions, the four-lepton event shapes 
observed by the ATLAS and CMS experiment \cite{Krasny:2013aca}.

\section{Summary and outlook}

We have briefly described the program \WINHAC\ and its
applications to some physics studies at the LHC. 
\WINHAC\ is an efficient and flexible Monte Carlo event generator 
for Drell--Yan processes at the LHC and other hadron colliders.
It features higher-order QED radiative corrections and is interfaced
with the \PYTHIA\ parton shower generator for QCD effects.
For the charged-current Drell--Yan processes it includes also the ${\cal O}(\alpha)$
electroweak corrections within the YFS exclusive exponentiation scheme.

The original \WINHAC\ program is written in the Fortran programming language.
Currently, its object-oriented versions in C++ are under development: 
for the charge-current DY processes called \WINHACpp~\cite{Sobol:2011zz}
and for the neutral-current ones called \ZINHAC~\cite{ZINHAC:MC}.
In the future we would like to interface these programs with our own
QCD parton shower generator, see
Refs.~\cite{Jadach:2011cr,Jadach:2012vs,Jadach:Ustron2013}.

\vspace{5mm}
\noindent
{\large\bf Acknowledgements}
\vspace{3mm}

\noindent
We thank D.\ Bardin, S.\ Bondarenko, F.\ Dydak, F.\ Fayette,
L.\ Kalinovskaya,  K.\ Sobol, K.\ Rejzner and A.\ Si\'odmok 
for the fruitful collaboration and the useful discussions.

\noindent


\end{document}